\documentclass[useAMS,usenatbib,onecolumn,preprint]{mn2e}
\usepackage{amsmath}
\usepackage{graphicx}
\usepackage{oldlfont}
\usepackage{longtable}
\usepackage{rotating}
\usepackage{amssymb}
\usepackage{subfloat}
\usepackage{float}
\usepackage[hang]{subfigure}

\def\jnl@style{\it}
\def\aaref@jnl#1{{\jnl@style#1}}

\def\aaref@jnl#1{{\jnl@style#1}}

\def\aj{\aaref@jnl{AJ}}                   
\def\apj{\aaref@jnl{ApJ}}                 
\def\apjl{\aaref@jnl{ApJ}}                
\def\apjs{\aaref@jnl{ApJS}}               
\def\apss{\aaref@jnl{Ap\&SS}}             
\def\aap{\aaref@jnl{A\&A}}                
\def\aapr{\aaref@jnl{A\&A~Rev.}}          
\def\aaps{\aaref@jnl{A\&AS}}              
\def\mnras{\aaref@jnl{MNRAS}}             
\def\prc{\aaref@jnl{Phys.~Rev.~C}}       
\def\prd{\aaref@jnl{Phys.~Rev.~D}}        
\def\prl{\aaref@jnl{Phys.~Rev.~Lett.}}    
\def\qjras{\aaref@jnl{QJRAS}}             
\def\skytel{\aaref@jnl{S\&T}}             
\def\ssr{\aaref@jnl{Space~Sci.~Rev.}}     
\def\zap{\aaref@jnl{ZAp}}                 
\def\nat{\aaref@jnl{Nature}}              
\def\aplett{\aaref@jnl{Astrophys.~Lett.}} 
\def\apspr{\aaref@jnl{Astrophys.~Space~Phys.~Res.}} 
\def\physrep{\aaref@jnl{Phys.~Rep.}}      
\def\physscr{\aaref@jnl{Phys.~Scr}}       
\def\commat{\aaref@jnl{Comm.~Math.~Phys.}}		
\def\science{\aaref@jnl{Science}}		
\def\cqg{\aaref@jnl{Classical Quant.~Grav.}}		
\def\jpcs{\aaref@jnl{JPCS}}					
\def\ijmpd{\aaref@jnl{Int.~J.~Mod.~Phys.~D}}			
\def\grg{\aaref@jnl{Gen.~Relat.~Gravit.}}		
\def\rpp{\aaref@jnl{Rep.~Prog.~Phys.}}		

\newcommand{\beq}{\begin{equation}}
\newcommand{\eeq}{\end{equation}}
\newcommand{\beqar}{\begin{eqnarray}}
\newcommand{\eeqar}{\end{eqnarray}}

\title[Coupled polar-axial magnetar oscillations] {Coupled polar-axial magnetar oscillations} 

\author[ A. Colaiuda and K. D. Kokkotas]
{ A.Colaiuda$^1$\thanks{colaiuda@tat.physik.uni-tuebingen.de} and K. D. Kokkotas$^{1,2}$\thanks{kostas.kokkotas@uni-tuebingen.de}\\
  $^1$Theoretical Astrophysics, University of T\"{u}bingen,  IAAT, Auf der Morgenstelle 10, T\"ubingen 72076, Germany\\
 $^2$Department of Physics, Aristotle University of Thessaloniki,
  Thessaloniki 54124, Greece }

\begin{document}


\maketitle
\begin{abstract}
We study  coupled axial and polar axisymmetric oscillations of a neutron star endowed with a strong magnetic field, having both poloidal and toroidal components.  The toroidal component of the magnetic field is driving the coupling between the polar and axial oscillations. The star is composed of  a fluid core as well as  a solid crust. Using a two dimensional general relativistic simulation and a magnetic field $B=10^{16}$G, we study the change in the polar and axial spectrum caused by the coupling. \textit{We find that the axial spectrum suffers a dramatic change in its nature, losing its continuum character}. In fact, we find that only the \textquoteleft edges' of the continua survive, generating a discrete spectrum. As a consequence the crustal frequencies, that in our previous simulation could be absorbed by the continua, if they were embedded inside it, are now long living oscillations.  They may lose their energy only in the very special case that they are in resonance with the  \textquoteleft edges' of the continua.

\end{abstract}


\section{Introduction}

In the last few years, a great attention was devoted to the interpretation of
the quasi periodic oscillations (QPOs) discovered in the tail of giant flares in Soft Gamma Repeaters (SGRs).  Those oscillations cover a wide range of frequencies going from a few Hz up to kHz. Until now only in two SGRs it was possible to observe clearly QPOs: in the SGR 1806-20 and in the SGR 1900+14 (see \cite{I2005}, \cite{Ws2006a} for a  review). In the SGR 1900+14  four frequencies have been detected: 28, 53, 84, and 155 Hz while in the SGR 1806-20 several QPOs have been discovered: 18, 26, 30, 92, 150, 625 and 1840 Hz. A recent reanalysis of the data for the SGR 1806-20  carried on by \cite{2011A&A...528A..45H} with a different method, shows the presence of additional QPOs at 16.9, 21.4, 36.4, 59.0 and 116.3 Hz.   

Understanding the nature of the QPOs will offer  an unique opportunity to investigate the properties of neutron stars and  to constrain the nuclear matter in neutron stars as well as the topology and strength of their magnetic field. For those reasons, since their discoveries, QPOs have been the subject of an intensive study. The first hypothesis, in order to explain QPOs, was to consider  pure shear modes of the crust, excited after the burst generated by giant flares. This idea proposed by \cite{1998ApJ...498L..45D} was investigate in several analytical and numerical works (see for example \cite{2007MNRAS.375..261S}, \cite{SA2007}). Those works evidenced that not all the lower QPOs could be explained as pure crustal frequencies. An alternative scenario has been then proposed by \cite{2006MNRAS.368L..35L} and \cite{2006MNRAS.371L..74G}, involving  global crust-core oscillations of the star. In particular,  \cite{2007MNRAS.377..159L} showed with a toy model that the QPOs spectrum should  be continuum, with turning points that were called \textquoteleft edges' of the continuum. Following this idea, a significant amount of  work has been devoted  in understanding global Alfv\'en oscillations in neutron star, using general relativistic models (\cite{2008MNRAS.385L...5S}, \cite{2009MNRAS.396.1441C}, \cite{2009MNRAS.397.1607C}) or Newtonian model (\cite{2008MNRAS.385.2069L}).  All those works made clear that global Alfv\'en modes in a pure fluid neutron star could not   explain the small gap between the lower frequencies observed in QPOs. 

A further step was then needed: to add a solid crust and to study the effect of its presence on the oscillations of the core ( see  \cite{2010JPhCS.222a2031K}). \cite{2011MNRAS.410.1036H}, using a non-relativistic model, showed that the presence of the crust strongly influences the continuum spectrum. In particular, they found that near the edges of the continua other discrete Alfv\'en modes appear.
This results was partially confirmed by \cite{2011MNRAS.410L..37G} and \cite{2012MNRAS.421.2054G}, using a non-linear relativistic method: the authors could find some frequencies in the gaps between two near continua, however they stated that those oscillations were quickly damped.
 In a linear relativistic simulation, 
\cite{2011MNRAS.414.3014C} confirmed the presence of those discrete Alfv\'en modes in the gap between two  continua and showed that the crustal modes could live long if they are located in those gaps.  The discrete Alfv\'en modes as well as the crustal modes in the gap between the first two continua could explain the somehow dense spectrum of the   observed lower  QPOs. However as pointed out in \cite{2012MNRAS.420.3035V}, it was still difficult to explain the higher frequencies as the 625Hz because at higher frequencies the continuum dominates, absorbing all the discrete frequencies, and in addition the continua  overlap to each other: in this way it is not possible to isolate the edges of the continuum and to have a unique interpretation for the 625Hz QPO.

A possibility to explain the high frequency QPOs came from the study of polar oscillations in magnetars. The work by   \cite{2009MNRAS.395.1163S} showed that the polar Alfv\'en spectrum has a discrete nature and, in addition, the lower polar Alfv\'en modes have frequencies  around few hundred Hz. However, it was not clear how this result could affect the axial spectrum.

In this work, we make an attempt to investigate the effect of the coupling between axial (torsional)  and polar oscillations  on the  spectrum. Our magnetar model is composed of a fluid core and a solid crust and is permeated by a magnetic field with strength $B=10^{16}$G which acquires both poloidal and toroidal components. The toroidal component is the coupling \textquoteleft pipeline' between the axial and polar part of the spectrum. 
By using a two dimensional relativistic, linear code, we find that the axial spectrum is strongly modified by the presence  of the coupling: its continuum feature is destroyed, leaving behind only its \textquoteleft edges'. In this new picture, both Alfv\'en  and crustal modes have a discrete nature and the latter ones, if out of resonance with the   \textquoteleft edges' of the continuum, could live long enough to be seen.

The paper is organized as follows: in Section 2 we present our background model, while in Section 3 we derive the perturbation equations and  test our code by comparing it with earlier results. In Section 4 we study a configuration where the coupling between axial and polar oscillations is active only on the axial part (i.e. we do not take into account the coupling to  the polar part). In Section 5 the full coupling between axial and polar oscillations is presented, including boundary conditions and initial data. The Section 6 and the Section 7 are devoted to comments and conclusions.

\section{Background Model}
We consider a non-rotating spherical symmetric star, whose metric is given by:
\begin{equation}
\label{metric}
ds^2=-e^{-2\Phi}dt^2+e^{2\Lambda}dr^2+r^2(d\theta^2+\sin^2 \theta d\phi^2)
\end{equation}
We restrict our model to   axisymmetric perturbations: all physical  quantities are not dependent on the $\phi$ coordinate.
The only non-zero term of the unperturbed four-velocity is $u^t$, i.e.   $u^\mu=(e^{-\Phi},0,0,0)$. 
 Finally, we do not consider the deformations induced on the star by the presence of a strong magnetic field, because the magnetic field has an energy $E_m$ which is a few orders of magnitude
smaller than the gravitational binding energy $E_g$. Typically, $E_m/E_g
\simeq 10^{-4}(B/10^{16}G)^2$ (see  \cite{2008MNRAS.385.2080C} and \cite{2008MNRAS.385..531H}  for numerical results about magnetars deformations).

We assume that the star is constituted of a perfect fluid in the ideal MHD approximation and of  a thin crust, that, as first approximation, can be described by a  shear velocity $v_S$:
\begin{equation}
\label{shear}
v_S=\frac{\mu}{\epsilon}=10^8\;cm\,s^{-1}
\end{equation}
where $\epsilon$ is the density and $\mu$ is the shear modulus. Note that the presence of nonuniform nuclear structure (the so called \textquoteleft pasta-phase') in the crust  could reduce the frequencies of the shear modes,  as it was recently suggested by \cite{2011MNRAS.417L..70S}. 

In this work, we neglect the superfluid component present both in the crust and in the core as well as the possible presence of proton superconductivity in the outer core (see  \cite{2011MNRAS.410..805G}), although we know that their appereance could strongly influence the features of the oscillations spectrum both in a qualitative way, changing its nature  from continuum to discrete, and in a quantitative way, shifting the frequencies (for more details see  \cite{2009MNRAS.396..894A}).

The magnetic field inside the star is described by the Grad-Shafranov equation:
\begin{equation}\label{GS}
a_1{''}e^{-2\Lambda}+\left(\Phi'-\Lambda' \right)e^{-2\Lambda}{a_1}'+\biggl(\zeta^2 e^{-2\Phi}-\frac{2}{r^2}\biggr)a_1=-4\pi j_1 \, \, .
\end{equation}
where $j_1$ and $a_1$ are respectively  the radial component of the four-current and the radial component of the magnetic field.
 Solving the equation (\ref{GS})  with the appropriate boundary conditions,  we get  the magnetic field configuration inside the star.  In our model, the magnetic field has both poloidal and toroidal components and it is given by:
\begin{eqnarray} \label{Bf}
 H^t=&0,\;\;\;\; &  H^r=\frac{e^{\Lambda} \cos \theta}{\sqrt{\pi}r^2}a_1, \\
 H^\theta=&-\frac{e^{-\Lambda} \sin \theta}{\sqrt{4\pi}}a_{1,r},\;\;\;\; & H^\phi=-\frac{e^{-\Phi} \sin \theta\zeta}{\sqrt{4\pi}}a_1,
\end{eqnarray}
where $\zeta$ is a parameter that describes the ratio of the toroidal magnetic field respect to the poloidal one, see \cite{2008MNRAS.385.2080C}. The interior magnetic field is matched with a pure poloidal magnetic field outside the star.
This choice implies that the toroidal magnetic field must be set equal to zero outside the star: in this way, a surface current  is created.

\section{Perturbation equations}

We consider both polar and axial perturbations, using the Cowling approximation, i.e $\delta g_{\mu \nu}=0$ so the components of the four-velocity $u^\mu$ are given by  \cite{2009MNRAS.395.1163S}:

\begin{eqnarray} \label{vf}
\delta u^t =&0\, , \\  \label{vf1}
 \delta u^r=&r^{-2}e^{-\Phi-\Lambda}\partial_t W(t,r,\theta)\, , \\ \label{vf2}
\delta u^\theta=&-r^{-2}e^{-\Phi}\partial_t  V(t,r,\theta)\, ,\\ \label{vf3}
\delta u^\phi=&e^{-\Phi-\Lambda}\partial_t {\cal Y} (t,r,\theta).
\end{eqnarray}

The polar perturbations involve  perturbations of the density and the pressure. For adiabatic perturbations and using the first law of the thermodynamics, the energy density and pressure variation could be written as:
\begin{equation}
\label{pert_dens}
\delta \epsilon= (\epsilon+p)\frac{\Delta n}{n}-\partial_r \epsilon \frac{W}{r^2} e^{-\Lambda}
\end{equation} 

\begin{equation}
\label{pert_press}
\delta p= \gamma p \frac{\Delta n}{n}-\partial_r p \frac{W}{r^2} e^{-\Lambda}
\end{equation} 
where $\gamma$ is the adiabatic constant and $\Delta n/n$ is the Lagrangian perturbations of the baryon number density.

With these assumptions, the linearized equations of motion are given by \cite{2008MNRAS.385L...5S}:
\begin{equation}\label{equa}
\begin{split}
(\epsilon+p+H^2)\delta u^\mu_{\;\;;\nu}u^\nu&=(\delta \epsilon+\delta p+2\delta H^\alpha H_\alpha)\ u^\mu_{\;\; ;\nu}\delta u^\nu\\&
+(u^\mu \delta u_\alpha+\delta^\mu u_\alpha)\biggl[H^\alpha H^\nu -g^{\alpha \nu}\biggl(p+\frac{1}{2}H^2\biggr)\biggr]_{;\nu}\\&
+h^\mu_{\alpha}\biggl[H^\alpha\delta H^\nu+\delta H^\alpha H^\nu-g^{\alpha \nu}(\delta p+H_\beta \delta H^\beta) \biggr]_{;\nu}\\&-h^\mu_\alpha\delta T^{\alpha \nu}_{\;\;\; ;\nu}
\end{split}
\end{equation}

where $h_{\mu \nu}=g_{\mu \nu}+u_\mu u_\nu$ and $\delta T^{\mu \nu}_{\;\; \; ; \nu}$ is the linearized shear stress tensor (see the Appendix  for the full derivation of the shear stress tensor).
The equation (\ref{equa}) leads to three partial differential equations (PDEs) for $\mu=r,\theta,\phi$, which have to be solved  with the appropriate boundary and initial conditions. 

\subsection{Comparison with earlier results}
\begin{figure}
  \centering
          {  \label{eig1}\includegraphics[width=0.4\textwidth]{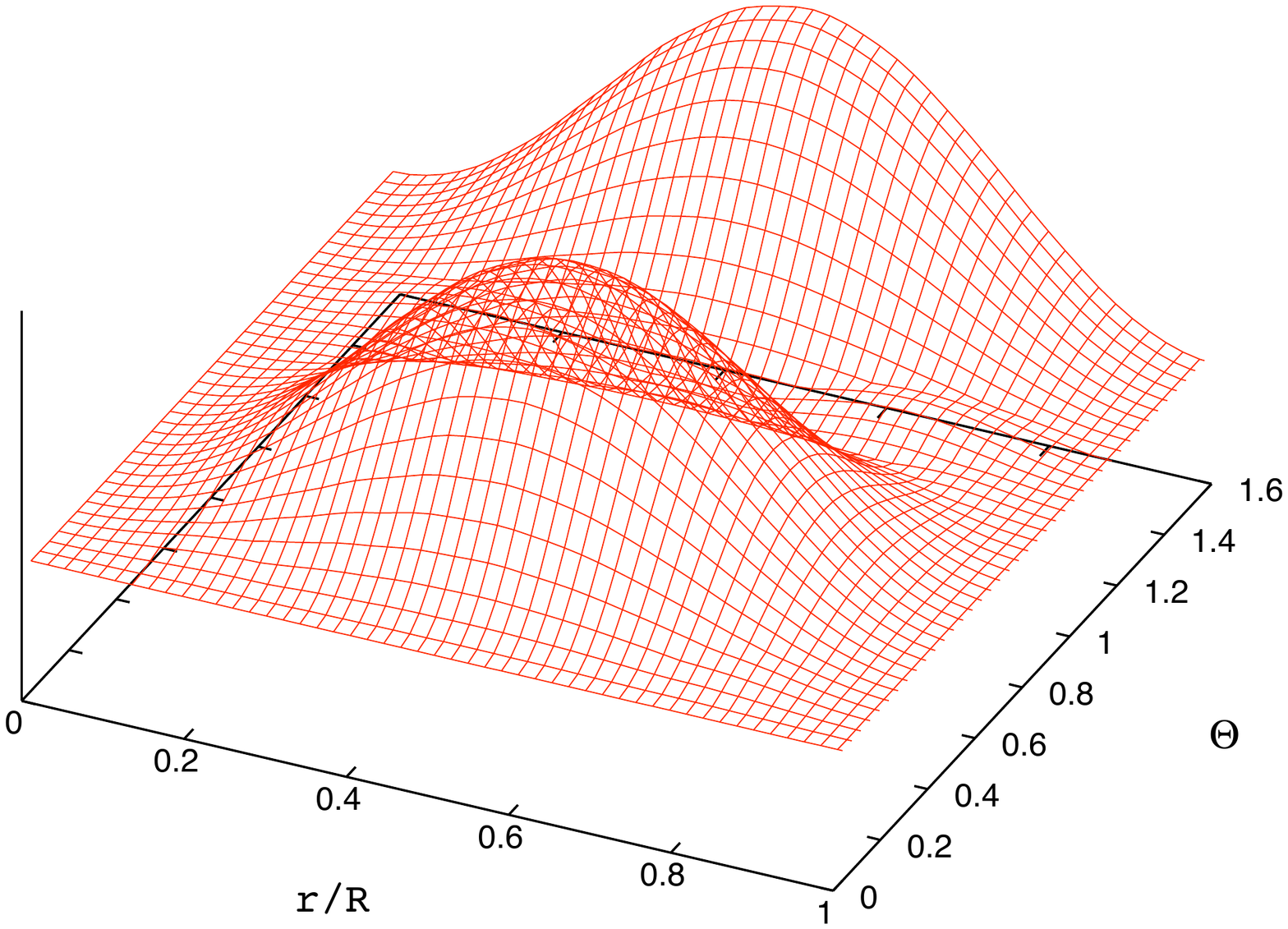}}      
{\label{eig2}\includegraphics[width=0.4\textwidth]{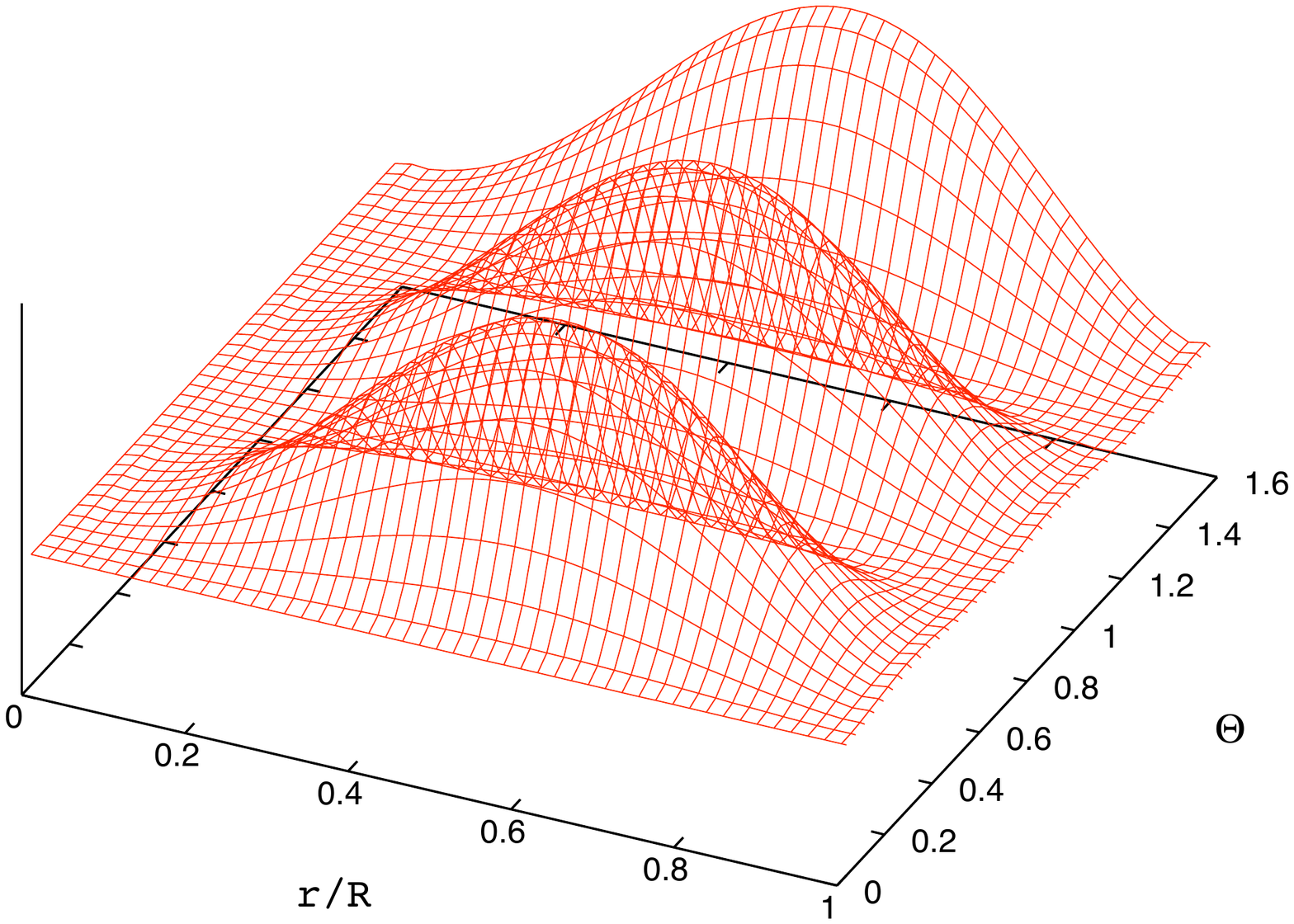}}
  \caption{The eigenfunction of the fundamental polar mode for the function $w(t,r,\theta)$ at 300Hz for $\ell=2$ on the left and of  the fundamental polar mode at 500Hz for $\ell=4$ on the right. Those plots are in good agreement, respectively, with the first plot  and  the third plot in figure 3  in \citep{2009MNRAS.395.1163S}.}
  \label{fig:eig1-2}
\end{figure}
In order to check our code for the polar oscillations, we implement the equation (22) of \cite{2009MNRAS.395.1163S} considering an axisymmetric non-rotating neutron star, permeated with a pure poloidal magnetic field. To be consistent with the model used in \cite{2009MNRAS.395.1163S}, the star is constituted of a pure fluid without the presence of a solid crust.
To get the code stable, we use an artificial viscosity of the second order. Note that in \cite{2009MNRAS.395.1163S} the artificial viscosity is of the fourth order. 
Despite this difference, our results are in good agreement with the ones found  by \cite{2009MNRAS.395.1163S}, as it can be seen from figure \ref{fig:eig1-2}, where the eigenfunctions  for the polar modes for $\ell=2$ and $\ell=4$ are plotted.
\vspace{-0.5cm}
\section{First configuration}

In order to investigate the coupling between polar and axial oscillations: we  should solve the system of equations described in (\ref{equa}) which involve  both the polar ($\mu=r,\,\theta$) and axial part ($\mu=\phi$), obtaining three coupled but different PDEs  for the functions ${\cal Y}, W $ and $V$. Each of those equations has the form: 
\begin{equation}
\label{eq_exp}
\partial_{tt} \mathbf{F} \mathit{(t,r,\theta)}=\nabla^2 \mathbf{F}+\mathit{coupling\;\; terms}
 \end{equation}  
where $F$ could be one of  the function ${\cal Y}$, $W$ and $V$ and the coupling terms involve the derivative both in space and time of the remaining two functions. The equations that we get are quite lengthy and require a significant amount of  computational work to be solved. In order to speed up our computational time  and as first attempt to understand the effect of the coupling, we choose to evolve the full equation (\ref{eq_exp}) only for the axial function ${\cal Y}$, while for  $V$ and $W$ we solve the equations (\ref{eq_exp}) without the presence of the coupling terms, setting them artificially equal to zero.

In this approximation, the equation (\ref{equa}) reduces for $\mu=r,\theta $ to the one found in \cite{2009MNRAS.395.1163S}. However, in addition to \cite{2009MNRAS.395.1163S} we consider the presence of a solid crust while  the magnetic field has  not only a poloidal component but also a toroidal one. As we explain later, this is needed in order to provide a  coupling between the polar and axial oscillations: notice that without the presence of toroidal field, there is no coupling between the classes of perturbations. An alternative way to have a consistent  coupling between the polar and the axial oscillations is to consider the deformation that the magnetic field induces on the shape of the star: however, this means to consider a perturbation of the metric given in (\ref{metric}), that will lead to a much more complicated structure of the final equation to solve and probably considerable weaker coupling.
Furthermore, there is a general belief that a combined poloidal-toroidal magnetic field could form a stable configuration (see \cite{2006A&A...450.1077B}).

The analytic form  of equation (\ref{equa}) for $\mu=\phi$ is:
\begin{equation}\label{equa1}
\begin{split}
A_{00}\frac{\partial^2 {\cal Y}}{\partial t^2}&=A_{20}\frac{\partial^2 {\cal Y}}{\partial r^2}
+A_{11}\frac{\partial^2 {\cal Y}}{\partial r \partial \theta}+A_{02}\frac{\partial^2 {\cal Y}}{\partial \theta ^2}
+A_{10}\frac{\partial {\cal Y}}{\partial r}+A_{01}\frac{\partial {\cal Y}}{\partial \theta}\\
&-\frac{\zeta \cot \theta e^{-2\Lambda-\Phi}}{\pi r^6}\biggl[ -\frac{a_1^2}{r}\biggl(\Phi^{\prime}+\Lambda^{\prime}+\frac{3}{r}\biggr)+\frac{a_1 a_1^{\prime}}{4}\biggl(\Phi^{\prime}+\Lambda^{\prime}+\frac{4}{r}\biggr) +\frac{1}{2}a_1^2\Lambda^{\prime}\Phi^{\prime}+ \frac{1}{2}a_1^{\prime 2}+\frac{1}{2}a_1^2\Phi^{\prime \prime}-\frac{1}{4}a_1^{\prime \prime}a_1\biggr]W\\&-\frac{\zeta \cot \theta e^{-2\Lambda-\Phi}}{\pi r^6}\biggl[ \frac{a_1^2}{r}\biggl(\Phi^{\prime}+\Lambda^{\prime}+\frac{4}{r}\biggr)-\frac{a_1 a_1^{\prime}}{2}\biggr]W_{,r}+\frac{\zeta  e^{-2\Lambda-\Phi}}{\pi r^6}\biggl[ \frac{a_1 a_1^{\prime}}{4}\biggl(\Phi^{\prime}+\frac{2}{r}\biggr) \biggr]W_{,\theta} +\frac{\zeta \cot \theta e^{-2\Lambda-\Phi}}{2\pi r^6}a_1^2 W_{,rr}
\\&-\frac{\zeta  e^{-2\Lambda-\Phi}}{\pi r^6}a_1 a_1^{\prime} W_{,r\theta}+\frac{\zeta \cot \theta e^{-3\Phi}}{2\pi r^6}a_1^2 W_{,tt}\\&+\frac{\zeta  e^{-\Lambda-\Phi}}{\pi r^6}a_1 a_1^{\prime}V_{,\theta \theta}-\frac{\zeta  e^{-\Lambda-\Phi}}{\pi r^6}a_1^2 V_{,\theta r}-\frac{\zeta  e^{-\Lambda-3\Phi}}{4\pi r^6}a_1 a_1^{\prime}V_{,tt}-\frac{\zeta \cot \theta  e^{-\Lambda-\Phi}}{\pi r^6}\biggl(\frac{3}{4}a_1 a_1^{\prime}-\frac{a_1^2}{r}\biggr)V_{,\theta }\\&-\frac{\zeta  e^{-\Lambda-\Phi}}{\pi r^6}\biggl(\frac{1}{2}a_1 a_1^{\prime}-\frac{a_1^2}{r}\biggr) V -\frac{\zeta  e^{-\Lambda-\Phi}}{2\pi r^6}a_1^2 V_{,r}.
\end{split}
\end{equation}
In this equation the coefficients $A_{00},\;A_{20},\; A_{11},\;A_{02},\;A_{01},\;A_{10} $ 
 depend 
 on the coordinates $r$ and $\theta$ but not on time. They are given by 
\begin{eqnarray}\label{att}
A_{00}&=&\biggl[\epsilon+p+\frac{a_1^2}{\pi r^4}\cos^2(\theta)+\frac{{a_1}'^{2}}{4\pi r^4}e^{-2\Lambda}\sin^2(\theta)\biggr]e^{-2(\Phi-\Lambda)} \, \, , \\
\label{a20}
A_{20}&=&\frac{a_1^2}{\pi r^4}\cos^2(\theta)+\mu \, \, , \\
\label{a11}
A_{11}&=&-\frac{a_1{a_1}{'}}{\pi r^4}\cos(\theta)\sin(\theta) \, \, , \\
\label{a02}
A_{02}&=&\frac{{a_1}'^{2}}{4\pi r^4}\sin^2(\theta)+\frac{\mu}{r^2}e^{2 \Lambda} \, \, ,\\
\label{a10}
A_{10}&=&(\Phi{ '}-\Lambda{ '})\frac{a_1^2}{\pi r^4}\cos^2(\theta)+
\frac{a_1{a_1}'^{}}{2\pi r^4}\sin^2(\theta)+\biggl[\mu{'}+\mu
\biggl(\frac{4}{r}-\Lambda{'}+\Phi{'}\biggr)\biggr] \, \, ,\\
\label{a01}
A_{01}&=&\biggl[\frac{a_1}{\pi r^4}\biggl(2\pi j_1-\frac{a_1}{r^2}\biggr)e^{2\Lambda}+3\frac{{a_1}'^{2}}{4\pi r^4}\biggr]\sin(\theta)\cos(\theta)+\frac{3}{r^2}\mu \cot(\theta)e^{2\Lambda} \, \, .
\end{eqnarray}

As we already pointed out,  the polar contribution in equation (\ref{equa1}) is always coupled with the toroidal field: in the absence of a toroidal field then the equation reduce to the one used for the study of  pure torsional oscillations \citep{2007MNRAS.375..261S}. Then equation  (\ref{equa1}) must be solved together  with equation (22) in \cite{2009MNRAS.395.1163S} for the polar functions $W$ and $V$.
 However, in \cite{2009MNRAS.395.1163S} the star consists of a pure fluid without crust: this we have  to calculate the crust contribution for the polar perturbations (see the Appendix for the full derivation).
The final form of the polar perturbation equations to be solved is:
\begin{eqnarray}
\label{polar1}
(\epsilon + p + H^2) e^{-\Phi} \delta u^{r}_{\ ,t} & = & - \left(\delta \epsilon + \delta p\right)\Phi' e^{-2\Lambda}+ H^r \delta H^\alpha_{\ ,\alpha} - e^{-2\Lambda}\delta p_{,r} + (\Lambda^{'} + \frac{2}{r})H^r\delta H^r   +  H^r \delta H^\alpha_{\ ,\alpha}           + H^\theta \delta H^r_{\ ,\theta} + H^\phi \delta H^r_{\ ,\phi} \nonumber \\
  & &    + [-2\Phi'e^{-2\Lambda}H_\theta + H^r_{\ ,\theta} -  2re^{-2\Lambda}H^\theta    + \cot\theta H^r - e^{-2\Lambda}H_{\theta,r}]\delta H^\theta - e^{-2\Lambda} H_\theta \delta H^\theta_{\ ,r}  \\
  &&  +-2\Phi'e^{-2\Lambda}H_\phi + H^r_{\ ,\phi} -2re^{-2\Lambda}\sin^2\theta H^\phi   - e^{-2\Lambda}H_{\phi,r}]\delta H^\phi  +e^{-2\Lambda}H_\phi\delta H^\phi_{\ ,r} + h^r_{\mu} \delta T^{\mu \nu}_{\;;\nu}, \nonumber
 \end{eqnarray}
\begin{eqnarray}
\label{polar2}
(\epsilon + p + H^2) e^{-\Phi} \delta u^{\theta}_{\ ,t} & = & -\frac{1}{r^2}\delta p_{,\theta} +[\bigl(\Phi^{'}+\Lambda^{'}+\frac{4}{r}\bigr)H^\theta+H^\theta_{,r}-\frac{1}{r^2}H_{r,\theta}]\delta H^r+ H^\theta \delta H^\alpha_{\ ,\alpha}-\frac{1}{r^2}H_r \delta H^r_{,\theta} \nonumber \\
  & &+\bigl[(\Phi^{'}+\frac{2}{r}\bigr)H^r+\cot \theta H^\theta\bigr] \delta H^\theta +H^r\delta H^\theta_{,\phi} \\ 
  & &    +\bigl [\frac{1}{r^2}(H_{\theta ,\phi}-H_{\phi , \theta})-\sin 2\theta H^\phi \bigr] \delta H^\phi-\frac{1}{r^2} H_{\phi} \delta H^\phi_{,\theta} \nonumber
   \end{eqnarray}

The equations above must be solved with the appropriate   boundary conditions listed in section \ref{boundary1}.

We want to point out that this type of configuration allows the transport of energy from the polar oscillations to the axial oscillations but not vice-versa. We will then expect to find the trace of the polar spectrum in the axial oscillations but no actual change in the polar spectrum. This is indeed what we find: the axial spectrum presents a significant excitation in the frequencies correspondent to the polar ones found in \cite{2009MNRAS.395.1163S}. Those frequencies do not present any significant excitation in the case of pure torsional oscillations, see \cite{2011MNRAS.414.3014C}. On the contrary, the polar spectrum does not show any change, as expected.   

This study demonstrates that, in order to understand how the polar oscillations are affected by  the coupling to the axial ones, we need to study the full problem, i.e. consider the equation (\ref{eq_exp}) also for the polar functions $W$ and $V$.

 \subsection{Boundary conditions}
\label{boundary1}

The equation (\ref{equa1}) must be solved with the appropriate boundary conditions for the axial function ${\cal Y}$ as well as for the polar functions $W $ and $V$. In order to simplify the boundary conditions on the surface, we use two new function $w(t,r,\theta)$ and $v(t,r,\theta)$ defined as (see \cite{2009MNRAS.395.1163S} ):

\begin{equation}
w=\epsilon W\;\;\; v=\epsilon V
\end{equation}

With this substitution, the equation (\ref{equa1}) becomes:

\begin{equation}\label{equa1b}
\begin{split}
A_{00}\frac{\partial^2 {\cal Y}}{\partial t^2}&=A_{20}\frac{\partial^2 {\cal Y}}{\partial r^2}
+A_{11}\frac{\partial^2 {\cal Y}}{\partial r \partial \theta}+A_{02}\frac{\partial^2 {\cal Y}}{\partial \theta ^2}
+A_{10}\frac{\partial {\cal Y}}{\partial r}+A_{01}\frac{\partial {\cal Y}}{\partial \theta}\\
&-\frac{\zeta \cot \theta e^{-2\Lambda-\Phi}}{\pi r^6}\biggl[ -\frac{a_1^2}{r}\biggl(\Phi^{\prime}+\Lambda^{\prime}+\frac{3}{r}\biggr)+\frac{a_1 a_1^{\prime}}{4}\biggl(\Phi^{\prime}+\Lambda^{\prime}+\frac{4}{r}\biggr) +\frac{1}{2}a_1^2\Lambda^{\prime}\Phi^{\prime}+ \frac{1}{2}a_1^{\prime 2}+\frac{1}{2}a_1^2\Phi^{\prime \prime}-\frac{1}{4}a_1^{\prime \prime}a_1\biggr]\frac{w}{\epsilon}\\&-\frac{\zeta \cot \theta e^{-2\Lambda-\Phi}}{\pi r^6}\biggl[ \frac{a_1^2}{r}\biggl(\Phi^{\prime}+\Lambda^{\prime}+\frac{4}{r}\biggr)-\frac{a_1 a_1^{\prime}}{2}\biggr]\biggl(\frac{w_{,r}}{\epsilon}-\frac{ w \epsilon_{,r}}{\epsilon^2}\biggr)+\frac{\zeta  e^{-2\Lambda-\Phi}}{\pi r^6}\biggl[ \frac{a_1 a_1^{\prime}}{4}\biggl(\Phi^{\prime}+\frac{2}{r}\biggr) \biggr]\frac{w_{,\theta}}{\epsilon} \\&+\frac{\zeta \cot \theta e^{-2\Lambda-\Phi}}{2\pi r^6}a_1^2 \biggl(\frac{w_{,rr}}{\epsilon}-\frac{2w_{,r}\epsilon_{,r}}{\epsilon^2}+\frac{2w \epsilon_{,r}^2}{\epsilon^3}-\frac{w\epsilon_{,rr}}{\epsilon^2}\biggr)
\\&-\frac{\zeta  e^{-2\Lambda-\Phi}}{\pi r^6}a_1 a_1^{\prime} \biggl(\frac{w_{,r\theta}}{\epsilon}-\frac{w_{,\theta}\epsilon_{,r}}{\epsilon^2}\biggr)+\frac{\zeta \cot \theta e^{-3\Phi}}{2\pi r^6}a_1^2 \frac{w_{,tt}}{\epsilon}\\&+\frac{\zeta  e^{-\Lambda-\Phi}}{\pi r^6}a_1 a_1^{\prime}\frac{v_{,\theta \theta}}{\epsilon}-\frac{\zeta  e^{-\Lambda-\Phi}}{\pi r^6}a_1^2 \biggl(\frac{v_{,\theta r}}{\epsilon}-\frac{v_{,\theta}\epsilon_{,r}}{\epsilon^2}\biggr)-\frac{\zeta  e^{-\Lambda-3\Phi}}{4\pi r^6}a_1 a_1^{\prime}\frac{v_{,tt}}{\epsilon}-\frac{\zeta \cot \theta  e^{-\Lambda-\Phi}}{\pi r^6}\biggl(\frac{3}{4}a_1 a_1^{\prime}-\frac{a_1^2}{r}\biggr)\frac{v_{,\theta }}{\epsilon}\\&-\frac{\zeta  e^{-\Lambda-\Phi}}{\pi r^6}\biggl(\frac{1}{2}a_1 a_1^{\prime}-\frac{a_1^2}{r}\biggr) \frac{v}{\epsilon} -\frac{\zeta  e^{-\Lambda-\Phi}}{2\pi r^6}a_1^2 \biggl(\frac{v_{,r}}{\epsilon}-\frac{v \epsilon_{,r}}{\epsilon^2}\biggr).
\end{split}
\end{equation}  

Using those new functions the boundary conditions are chosen as following: 

\begin{itemize}
  \item at the centre we require the  regularity of the equations:
\begin{equation}
\label{regularity}
{\cal Y}(r=0)=0, \;\;\; w(r=0)=0, \;\;\; v(r=0)=0\,,
\end{equation} 

 \item at the surface, we use the zero traction condition:
 \begin{equation}
\label{surface}
\partial_r {\cal Y}=0,\;\;\; \partial_r v(r=R)=0,\;\; \partial_r w(r=R)=0\,,
\end{equation} 
   
   \item on the magnetic axis at $\theta=0$, we require axisymmetry:
   \begin{equation}
\label{eq_l2}
\partial_\theta {\cal Y}=0,\;\;\; \partial_\theta w=0,\;\;\; \partial_\theta v=0\,.
\end{equation}

 \item At the equatorial plane, we require equatorial plane symmetry  at $\theta=\pi/2$:
   \begin{equation}
\label{eq_l22}
\partial_\theta {\cal Y} =0,\;\;\; \partial_\theta w=0,\;\;\; \partial_\theta v=0\,,
\end{equation}
or antisymmetry:
  \begin{equation}
\label{eq_l3}
{\cal Y}=0,\;\;\; w=0,\;\;\; v=0\,,
\end{equation} 

\item at  the crust-core interface,  we require the continuity of the functions:
\begin{equation}
\label{crust-core0}
\partial_r {\cal Y}=\biggl[1+\frac{v_s}{v_A}\biggr]\partial_r {\cal Y}\,,
\end{equation}
\begin{equation}
\label{crust-core1}
\partial_r v=\biggl[1+\frac{v_s}{v_A}\biggr]\partial_r v\, ,
\end{equation}
\begin{equation}
\label{crust-core2}
\partial_r w=\biggl[1+\frac{v_s}{v_A}\biggr]\partial_r w\,.
\end{equation}
\end{itemize} 

where $v_s$, the shear velocity, and $v_A$,  the Alfv\'en velocity, are respectively given by:
\begin{equation}
\label{vel_s_a}
v_s=\biggr(\frac{\mu}{\epsilon}\biggl)^{1/2}\simeq 10^{18} cm/s, \;\;\;\; v_A=\frac{H}{\sqrt{ \epsilon}}
\end{equation}

where $\mu$ is the shear modulus.

\section{Polar and axial oscillations: coupling}


We now derive explicitly the equations for the polar function $W$ and $V$, including the presence of the crust presented in   equations (\ref{Tr}) and (\ref{Tth}) and the coupling with the axial function $\cal {Y}$. Substituting the equations (\ref{vf1}) and (\ref{vf2}) in equations (\ref{polar1}) and (\ref{polar2}) we get:

\begin{equation}\label{polar_W}
\partial_{tt} w(\tilde{A_{00}}\tilde{A_{11}}-\tilde{A_{10}}\tilde{A_{01}})=F_v\tilde{A_{11}}-F_w\tilde{A_{01}}+A_{y_{00}}\partial_{tt} {\cal Y}+A_{y_{01}}\partial_\theta {\cal{Y}}+A_{y_{10}}\partial_r \cal{Y}
\end{equation} 

\begin{equation}\label{polar_V}
\partial_{tt} v(\tilde{A_{00}}\tilde{A_{11}}-\tilde{A_{10}}\tilde{A_{01}})=F_v\tilde{A_{00}}-F_w\tilde{A_{10}}+\tilde{A_{y_{00}}}\partial_{tt} {\cal Y}+\tilde{A_{y_{01}}}\partial_\theta {\cal{Y}}+\tilde{A_{y_{10}}}\partial_r \cal{Y}
\end{equation} 

where all the coefficients are function only of the coordinates $r,\,\theta$:

\begin{eqnarray} \label{coeff_coupling}
\tilde{A_{00}}=&(\epsilon+p+H^\theta H_\theta+H^\phi H_\phi)r^{-2}e^{-2\Phi-\Lambda}\, ,\\
\tilde{A_{01}}=&r^{-2}H^r H_\theta \, ,\\
\tilde{A_{10}}=&-r^{-2}H^\theta H_r \, ,\\
\tilde{A_{11}}=&-(\epsilon+p+H^r H_r+H^\phi H_\phi)r^{-2}e^{-2\Phi} \, ,\\
\tilde{A_{y_{00}}}=&-\zeta a_1 \cot{\theta}\, r^{-2} e^{-\Lambda} (\tilde{A_{11}} a_1^{'}-\tilde{A_{01}}a_1) \, ,\\
A_{y_{00}}=&-\zeta a_1  \cot{\theta}\, r^{-2 }e^{-\Lambda} (\tilde{A_{00}} a_1^{'}-\tilde{A_{10}}a_1) \, ,\\
\tilde{A_{y_{01}}}=&2{\zeta  a_1^{2} \sin{\theta}\cos{\theta}}{e^{-\Lambda}} \tilde{A_{11}} \, ,\\
A_{y_{01}}=&4{\zeta a_1^2 \sin{\theta}\cos{\theta}}{r^{-2} e^{-3\Lambda}} \tilde{A_{00}} \, ,\\
\tilde{A_{y_{10}}}=&2{\zeta  a_1^{2} \sin{\theta}\cos{\theta}}{e^{-\Lambda}} \tilde{A_{11}} \, ,\\
A_{y_{10}}=&4{\zeta a_1^2 \sin{\theta}\cos{\theta}}{r^{-2} e^{-3\Lambda}} \tilde{A_{00}} \, ,
\end{eqnarray}
while the two expressions ${\cal F}_W$ and ${\cal F}_V$ are (\cite{2009MNRAS.395.1163S}):
\begin{eqnarray}
 {\cal F}_W 
     &=& - \left(\delta \epsilon + \delta p\right)\Phi' e^{-2\Lambda} - e^{-2\Lambda}\delta p_{,r}
         + \left(\Lambda' + \frac{2}{r}\right)H^r\delta H^r
         + H^r \left(\delta H^r_{\ ,r} + \delta H^\theta_{\ ,\theta} + \delta H^\phi_{\ ,\phi}\right)
         + H^\theta \delta H^r_{\ ,\theta} + H^\phi \delta H^r_{\ ,\phi} \nonumber \\
     &&  + \left[-2\Phi'e^{-2\Lambda}H_\theta + H^r_{\ ,\theta} - 2re^{-2\Lambda}H^\theta
         + \cot\theta H^r - e^{-2\Lambda}H_{\theta,r}\right]\delta H^\theta
         - e^{-2\Lambda} H_\theta \delta H^\theta_{\ ,r} \nonumber \\
     &&  + \left[-2\Phi'e^{-2\Lambda}H_\phi + H^r_{\ ,\phi} -2re^{-2\Lambda}\sin^2\theta H^\phi
         - e^{-2\Lambda}H_{\phi,r}\right]\delta H^\phi - e^{-2\Lambda}H_\phi\delta H^\phi_{\ ,r}, \\
 {\cal F}_V 
     &=& - \frac{1}{r^2}\delta p_{,\theta} + \left[\left(\Phi' + \Lambda' + \frac{4}{r}\right)H^\theta
         + H^\theta_{\ ,r} - \frac{1}{r^2}H_{r,\theta}\right]\delta H^r
         + H^\theta \left(\delta H^r_{\ ,r} + \delta H^\theta_{\ ,\theta} + \delta H^\phi_{\ ,\phi}\right)
         - \frac{1}{r^2}H_r \delta H^r_{\ ,\theta} \nonumber \\
     &&  + \left[\left(\Phi' + \frac{2}{r}\right)H^r + \cot\theta H^\theta\right]\delta H^\theta
         + H^r \delta H^\theta_{\ ,r} + H^\phi \delta H^\theta_{\ ,\phi} \nonumber \\
     &&  + \left[\frac{1}{r^2}\left(H_{\theta,\phi} - H_{\phi,\theta}\right)
         - 2\sin\theta\cos\theta H^\phi\right]\delta H^\phi - \frac{1}{r^2}H_\phi \delta H^\phi_{\ ,\theta}.
\end{eqnarray}

Note that, as for the previous configuration, the coupling between the axial and the polar functions is always determined by the toroidal component of the magnetic field. 

\subsection{Numerical method}
We adopted the same equation of state (EOS) used in  \cite{2009MNRAS.395.1163S}, i.e a polytropic  EOS that can reproduce quite well the tabulated data for the realistic EOS A \citep{1971NuPhA.178..123P}. For more details about this EOS see section 4.1 in \cite{2009MNRAS.395.1163S}. We focus on a stellar model with a mass $M=1.4M_\odot$ and a radius $R=10.35$ km  while the value for the   magnetic field strength that we assume is $B_\mu= 10^{16}$ Gauss on the pole. This magnetic field is stronger than the magnetic field that we  used in our previous simulations. However, as was already pointed out by \cite{2009MNRAS.395.1163S}, a magnetic field of this strength is needed in order to get a cleaner imprint of the polar frequencies. We will explain the implications of such strong magnetic field  on the spectrum later.

In all the simulations that we present  the parameter $\zeta$ is set to $\zeta=0.1$. We perform also a simulation with $\zeta=0.2$ and we find that the energy transfer between the polar and the axial system becomes, as expected, more efficient.

We used a numerical  grid in the plane $(r,\theta)$ $150 \times 50$,  varying 
$\theta$ from $0$ to $\pi/2$.  The accuracy of the code was tested, performing a simulation with a $200\times50$ grid: the results show that the frequencies are not  influenced significantly by a change in the number of grid points. The stability of the scheme was also checked: our simulation lasted few seconds and the use of artificial viscosity does not damp significantly the oscillations.
The base of crust ( $r_{\rm crust}$ ) is taken at $\epsilon=2.4\times 10^{14}$ $g/cm^3$, according to the crust model by  \cite{1973NuPhA.207..298N} (in the following NV).  Core and crust are coupled via the interface conditions (\ref{crust-core0})-(\ref{crust-core2}).

As initial conditions, considering the requirement of regularity at the centre of the star, we use:
\begin{eqnarray} \label{initial_condition}
w(r,\theta)&=&\epsilon r^{\ell+1}\sin \theta P_\ell(\cos \theta)\,,\\
v(r,\theta)&=&-\frac{\epsilon}{\ell} r^{\ell}\sin \theta \partial_\theta P_\ell(\cos \theta)\,,\\
{\cal Y}(r,\theta)&=&\epsilon r^{\ell+1}\sin \theta P_\ell(\cos \theta)\,.
\end{eqnarray}

\section{Results \& comments}

\subsection{On the nature of the spectra}
We are interested to understand not only how the perturbations propagate in the star but also to determine the changes that the coupling between polar and axial oscillations brings on the stellar spectrum. In particular, we want to investigate if the nature of the spectrum (discrete for polar modes and continuous for axial ones) is preserved or if it changes substantially.
In addition, we want to study how efficiently the oscillations are propagating when just one type of perturbation is excited.  
  
For this reason, we  performed different runs, exciting initially just the polar oscillations and looking for the propagation of the perturbation on the axial part.  We notice that the axial part begins to oscillate immediately  after the onset of the perturbation. This could be easily understood if we consider what is the frequencies range of both polar and axial oscillations: the polar oscillations have frequencies going from 300 Hz to kHz, while the lower axial frequency has been found  around $~16-20$ Hz, depending on the EOS used. This means that, while the axial oscillations need a longer evolution in order to be  clearly detected, the polar oscillations could appear after a much smaller amount of evolution time. However,  the coupling between the two systems is so efficient, even for relatively small values of the  toroidal magnetic field, that  the energy can flow quickly from one system to the other.  In this way,  the axial oscillations have been  excited significantly even if the initial perturbation was only on the polar part of the equations.  However, the axial oscillations have been influenced by  the presence of the coupling to the polar ones in a very dramatic way. 

\begin{figure}
\begin{center}
\includegraphics[width=0.7\textwidth]{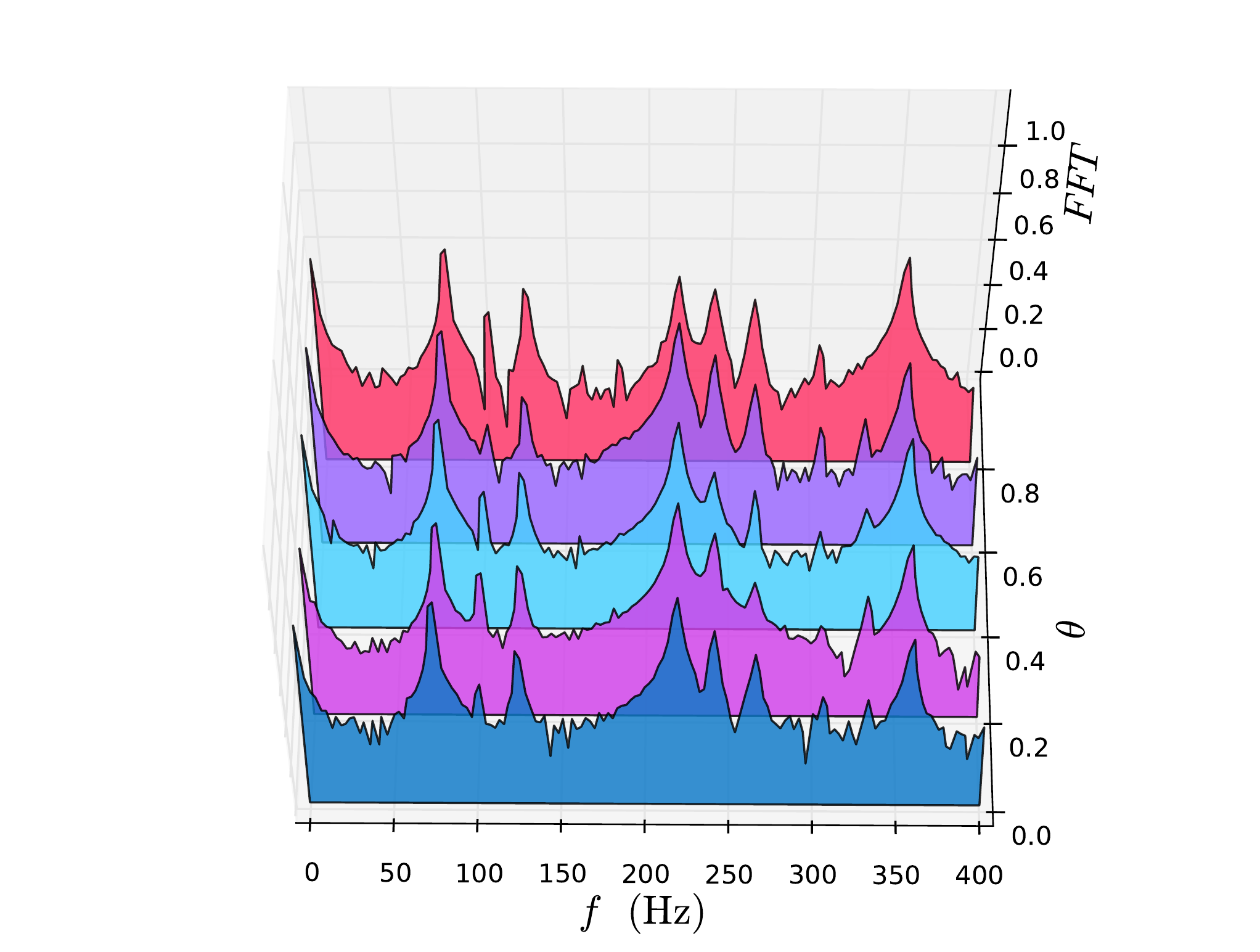}
\caption{
The FFT of the function ${\cal{Y}}$ taken at a fixed radius and varying the angular position of the observing point, i.e. the value of $\theta$. It is evident that the peaks are independent from the observer's position. In addition, we can see a particular excitation in the axial spectrum at high frequencies (around 200Hz), probably due to the coupling with the polar spectrum.}%
\label{disc}
\end{center}
\end{figure}

We know from previous works  (\cite{2006MNRAS.368L..35L}, \cite{2006MNRAS.371L..74G}, \cite{2008MNRAS.385L...5S}, \cite{2009MNRAS.396.1441C}, \cite{2009MNRAS.397.1607C}, \cite{2011MNRAS.410.1036H}, \cite{2011MNRAS.410L..37G}, \cite{2011MNRAS.414.3014C}, \cite{2012MNRAS.421.2054G} ) that the spectrum of the axial oscillations has a continuum component generated by the presence of a strong magnetic field inside the star. \textit{The presence of the coupling between axial and polar oscillations changes this picture and the continuum reduces to a discrete spectrum}. We observe this change in the following way: we  choose a point with certain ($r,\theta$) coordinates  inside the star, then we take the  FFT of this point, getting a certain frequency that corresponds to a certain mode. Then, by fixing the value of $r$, we change the value of $\theta$, i.e. we choose a  different phase, and we perform again a FFT of the point, getting a new frequency, and so on, scanning all the $\theta$ range. As it can be seen in figure \ref{disc}, the frequencies that we get fixing $r$ and varying the value of $\theta$ do not change, confirming their discrete nature. However, comparing those new results with the ones that we found in \cite{2011MNRAS.414.3014C}, we noticed that those \textquoteleft new' discrete frequencies correspond to the edges of the continuum that we found previously.

It is then important to examine if the new nature of the torsional oscillations spectrum  is consequence of the interface condition between crust and core and of the coupling of the polar and axial oscillation or it is  a consequence of the coupling alone. In order to understand this, we solve the equations (\ref{equa1}), (\ref{polar_V}) and (\ref{polar_W}) for a pure fluid star without  a solid crust. We find that the new discrete character of the torsional oscillations is preserved even in this alternative scenario, showing that the discrete nature of the spectrum is a consequence of the coupling between polar and axial oscillations. Since the coupling happens through the toroidal component of the magnetic field, \textit{it can be stated that the magnetic field configuration has a remarkable role in determining the final nature of the spectrum of the torsional oscillation}. In fact, we find that even for small values of the toroidal magnetic field, the coupling between axial and polar oscillations is already  effective enough to change the discrete nature of the torsional oscillations.

In addition, the axial spectrum presents  strong excitations at high frequencies, probably due to the coupling with the polar spectrum: in fact, as we already pointed out,   the first lower polar frequency is located at few hundred Herz.  The strong excitation at high frequencies for the function ${\cal {Y}}$ can   be seen in figure \ref{disc}. This excitation was not present in the case of pure axial modes (\cite{2011MNRAS.414.3014C}). 

\subsection{Gravitational waves}
The polar spectrum present variations too. In particular, we can see the appearance of lower frequencies that are imprints of the coupling with the axial oscillations. These frequencies correspond to the discrete ones of the axial spectrum.
One could think that this last feature of the polar spectrum   could have an important effect  in the possibility of detecting gravitational waves from magnetars. In fact, we know from the works by  \cite{2011MNRAS.tmp.1405L}, \cite{2011ApJ...735L..20L}, \cite{2011ApJ...736L...6C},  \cite{2012PhRvD..85b4030Z} that the possibility of detecting gravitational waves from torsional oscillations or hydromagnetic instability in  magnetars is quite low. However, if density perturbations are excited in the polar spectrum by the coupling with the torsional one, then the possibility of observing gravitational waves could increase significantly. For this reason, we study the density perturbations that have been generated by the coupling between the axial and polar oscillations through the fromula:
\begin{equation}
\label{per_dens}
\delta \epsilon=-\frac{(p+\epsilon)}{r^2 } \biggl(-\frac{ w_{,r}}{\epsilon e^\Lambda}+\frac{ \epsilon_{,r}w}{\epsilon^2 e^\Lambda}+ \frac{1}{r^2}\frac{ v_{,\theta}}{\epsilon}+\frac{\cot \theta v}{r^2 \epsilon} \biggr)-\frac{1}{r^2 e^{\Lambda}}\frac{w \epsilon_{,r}} {\epsilon }.
\end{equation}
\begin{figure}
\begin{center}
{\label{comp}\includegraphics[width=0.45\textwidth]{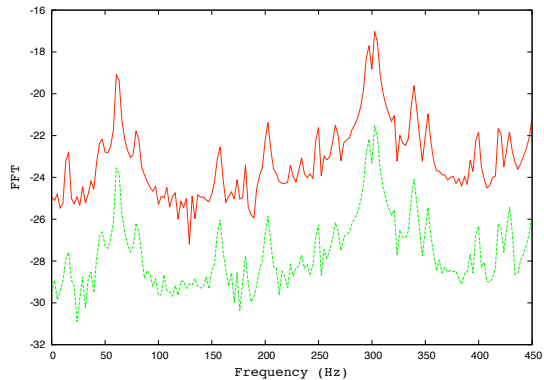}}
\caption{
 The FFT of the polar function  $w(t,r,\theta)$ (red line) and the corresponding density perturbations $\delta \epsilon$ (green line). The amplitude of the density perturbation is several order of magnitude smaller   than the one of the polar function $w(t,r\theta)$: this implies that the density perturbations are not excited to a significant level and, as consequence, the possibility of detecting gravitational waves is quite low.
}%
\label{axial1}
\end{center}
\end{figure}


The results are shown in figure \ref{axial1}: the density perturbations are several orders of magnitude lower than the normal axial perturbations. We  then conclude that  density perturbations are not excited in a significant level and the possibility of detecting gravitational waves remains  quite low.

\subsection{On the nature of some special modes}
It is worth to notice that in \cite{2011MNRAS.414.3014C}, two families of  discrete frequencies have been found: we called them \textquoteleft crustal modes' and \textquoteleft discrete Alfv\'en modes': while the former modes are shear axial modes that can be recovered even in absence of magnetic field, the latter are a consequence of the presence of the solid crust and of its coupling  with a fluid core and they are directly proportional to the magnetic field strength.  In addition, in \cite{2011MNRAS.414.3014C}, we found that if a crustal mode is embedded in the continuum, then its energy  will be quickly damped and eventually absorbed by the continuum. In the configuration that we study in this paper the continuum disappears and then the crustal modes can be effectively damped only if they are in resonance with the discrete frequencies of the continuum. The speed of the damping process depends on the strength of the magnetic field: for high values of the magnetic field, the crustal frequencies, if in resonance with the discrete modes of the continuum, are absorbed considerably faster.  

Regarding the \textquoteleft discrete Alfv\'en frequencies', it is difficult, in this new model, to distinguish between the edges of the continuum and the \textquoteleft discrete Alfv\'en frequecies' since both of them have now a discrete nature and both of them are proportional to the magnetic field. 

Concerning the identification of the QPOs, we try to solve the puzzle of the 625 Hz QPO that has been reported in \cite{SW2005}, 
\cite{2007AdSpR..40.1446W}. In our previous work we could not explain this long living QPO because, at high frequencies, the edges of  the continua present in absence of coupling could not be isolated anymore since the various continua start to overlap with each other, and the crustal mode are quickly absorbed from the continua. In this new model, the torsional spectrum is discrete but the spectrum of the oscillations at high frequencies   are too dense to allow an univocal explanation for the 625 Hz QPO.  
This QPO could still be explained as a higher  polar mode.


\section{Conclusion}

In this paper we study the coupling between the axial and the polar type of oscillations in a strongly magnetised neutron star, composed of a fluid core and a solid crust and permeated by a magnetic field with both a toroidal and a poloidal component. This mixed magnetic field configuration  is though to be stable (see \cite{2006A&A...450.1077B}, \cite{2011ApJ...735L..20L}, \cite{2011ApJ...736L...6C} and references therein).
Following the work by  \citep{2009MNRAS.395.1163S}  (for the polar part) and \cite{2011MNRAS.414.3014C} (for the axial part), we derive a coupled system of equations that involves both the polar perturbation functions $W$ and $V$ and the axial function $\cal {Y}$. The equations are coupled through the toroidal component of the magnetic field. After solving those equations with the appropriate boundary conditions and  various sets of initial data, we obtain the axial and polar spectrum. 

However, we found that both the spectra are strongly influenced by the coupling: the axial spectrum, that in absence of coupling had a discrete nature with the lower frequency located around a few hundred Hertz, preserves its discrete nature but some very low frequencies appear in the 20-300 Hz range. Those lower frequencies seems to be strictly related to the energy transfer between the axial and the polar oscillations via toroidal magnetic field: the lower frequencies found in the axial spectrum are the same found in the polar one, with a tiny shift of 1 or 2 Hz. However, the most relevant change is the one that concerns the axial spectrum: examining the FFT amplitude at various points in the star by varying the phase, \textit{we observe that the axial spectrum has a discrete nature}. In fact, we can observe only the edges of the continuum that we found in \cite{2011MNRAS.414.3014C}: those edges are still Alfv\'en modes, in fact they scale with the strength of the magnetic field and their restore force is the magnetic force. 
\textit{Due to the presence of the crust in our model, also \textquoteleft crustal modes' appear in the spectrum}. In the axial case, they are not anymore absorbed by the continuum so they are long-living modes. The only case in which a damping of the crustal modes  can be observed is if they are in resonance with the edges of the continuum. In this case the energy leaks from the crust to the core with a velocity that depends on the strength of the magnetic field. The efficiency of the  process depends on the  magnetic field strength. Finally, we made an attempt  to resolve the puzzle of the 625 Hz QPO: we find that in our model, although the spectrum is discrete, at higher frequencies it becomes  too dense to allow an unambiguous interpretation of this high frequency mode.

In the future we would like to investigate the effect of superconductivity on the oscillations spectrum as well as non-axisymmetric oscillations. Note that a first study of non-axisymmetric  oscillations in a magnetised neutron star was made by \cite{2010MNRAS.405..318L}, \cite{2011MNRAS.412.1730L} (with a Newtonian model) and by \cite{2011JPhCS.314a2081S} (with a general  relativistic model): they both found that the spectrum of non-axisymmetric oscillations has also a discrete nature.

\section*{Acknowledgements}
We thank Y. Levin, N. Stergioulas and M. Gabler for fruitful discussions and E. Gaertig for providing us his mode recycling routine. This work was supported by the German Science Council (DFG) via SFB/TR7.

\section*{Appendix: The Shear Stress Tensor}
\label{sec_stress}

The linearized shear stress tensor $\delta S_{\mu \nu}$  is defined as:
\begin{equation}
\label{stress_T}
\delta T^{\mu \nu}=-2\mu \delta S^{\mu \nu}
\end{equation}
and it can be calculated from the formula:
\begin{equation}
\label{stress_S}
\delta \sigma^{\mu \nu}=e ^{-\Phi}\delta S^{\mu \nu}_{\;,t}
\end{equation}
Here  $\sigma_{\mu \nu}$ is defined as:
\begin{equation}
\label{stress_sigma0}
\sigma_{\mu \nu}=\frac{1}{2}(u_{\mu;\alpha}P^\alpha_\nu+u_{\nu;\alpha}P^\alpha_{\mu})-\frac{1}{3}P_{\mu \nu}u^\beta_{\;\beta}
\end{equation}
with $P_{\mu \nu}=g_{\mu \nu}+u_\mu u_\nu$. The linearized perturbation of  $\sigma_{\mu \nu}$ leads to
\begin{equation}
\label{stress_sigma}
\delta \sigma_{\mu \nu}=\frac{1}{2}(\delta u_{\mu;\alpha}P^\alpha_\nu+ u_{\mu;\alpha}\delta P^\alpha_\nu+\delta u_{\nu;\alpha}P^\alpha_{\mu}+u_{\nu;\alpha}\delta P^\alpha_{\mu})-\frac{1}{3}\delta P_{\mu \nu}u^\beta_{;\beta}-\frac{1}{3} P_{\mu \nu}\delta u^\beta_{\; ;\beta}
\end{equation}

The non-zero components of the  tensor $\delta \sigma_{\mu \nu}$ (\ref{stress_sigma}) are:
\begin{eqnarray}
\label{sigma_rr}
\delta \sigma_{rr}&=&\frac{e^{2\Lambda-\Phi}}{3r^2}\biggl(2e^{-\Lambda}W_{,tr}-6\frac{e^{-\Lambda}}{r}W_{,t}+V_{,t\theta}+\frac{\cos \theta}{\sin{\theta}}V_{,t}\biggr)\, ,\\
\label{sigma_thr}
\delta \sigma_{r\theta}&=&-\frac{e^{-\Phi}}{2}\biggl(V_{,tr}-\frac{2}{r}V_{,t}-\frac{e^{\Lambda}}{r^2}W_{,t\theta}\biggr)\, ,\\
\label{sigma_phr}
\delta \sigma_{r\phi}&=&\frac{1}{2}r^2e^{-\Phi}\sin{\theta}^2 Y_{,t r}\, ,\\
\label{sigma_thth}
\delta \sigma_{\theta\theta}&=&-\frac{e^{-\Phi}}{3}\biggl(e^{-\Lambda}W_{,tr}-\frac{3e^{-\Lambda}}{r}W_{,t}+2V_{,t \theta }-\frac{\cos \theta}{\sin \theta}V_{,t}\biggr)\, ,\\
\label{sigma_thph}
\delta \sigma_{\theta\phi}&=&\frac{1}{2} r^2e^{-\Phi}\sin{\theta}^2 Y_{,t \theta}\, ,\\
\label{sigma_phph}
\delta \sigma_{\phi \phi}&=&\frac{e^{-\Phi}}{3}\sin^2 \theta\biggl(e^{-\Lambda}W_{,tr}+\frac{3e^{-\Lambda}}{r}W_{,t}+V_{,t \theta }-2\frac{\cos \theta}{\sin \theta}V_{,t}\biggr)\, .
\end{eqnarray}

Using the above equations, we can easily calculate the shear stress  tensor $\delta T^{\mu \nu}$ from equation (\ref{stress_T}). Once the shear stress tensor is known, we can calculate the shear energy contribution present in equation (\ref{equa}) for the polar part (i.e. for $\mu=r$ and $\mu=\theta$). These contributions are given by:

\begin{equation}\label{Tr}
\begin{split}
h^{r}_{\alpha}\delta T^{\alpha \nu}_{\;\;\; ;\nu}=&-\frac{4}{3} \frac{\mu}{e^{2\Phi+3\Lambda}r^3}W_{,rr}- \frac{\mu}{e^{2\Phi+\Lambda}r^4}W_{,\theta \theta}+\frac{4}{3} \frac{\mu}{e^{2\Phi+3\Lambda}r^2}\biggl(\Phi^{\prime}+ \Lambda^{\prime}+\frac{2}{r} \biggr)W_{,r}-4 \frac{\mu}{e^{2\Phi-3\Lambda}r^3}\biggl(\Phi^{\prime}+\Lambda^{\prime}\biggr)W+\\& \frac{\mu}{e^{2\Phi+2\Lambda}r^2}V_{,\theta r}+ \frac{2}{3} \frac{\mu}{e^{2\Phi+2\Lambda}r^2}\biggl(\Phi^{\prime}-\frac{4}{r} \biggr)V_{,\theta}+\frac{1}{3} \frac{\mu \cot \theta}{e^{2\Phi+2\Lambda}r^2}V_{,r}+\frac{2}{3} \frac{\mu \cot \theta}{e^{2\Phi+2\Lambda}r^2}\biggl(\Phi^{\prime}-\frac{4}{r} \biggr)V,
\end{split}
\end{equation}

\begin{equation}\label{Tth}
\begin{split}
h^{\theta}_{\alpha}\delta T^{\alpha \nu}_{\;\;\; ;\nu}=& \frac{4 \mu}{3 e^{2\Phi}r^2}V_{,rr}+\frac{4 \mu}{3 e^{2\Phi}r^4}V_{,\theta \theta}+\frac{4}{3}\frac{\mu \cot \theta}{r^4}V_{,\theta}-\frac{e^{-2(\Phi+\Lambda)}\mu}{r^2}\biggl(\Phi^{\prime}+\Lambda^{\prime}\biggr)V_{,r}+ \\& \frac{2e^{-2(\Phi+\Lambda)}\mu}{r^3}\biggl(\Phi^{\prime}+\Lambda^{\prime}+\frac{1}{r}-\frac{2 e^{2\Lambda}}{3 r}\biggr)V-\frac{4 \mu e^{-2\Phi}\cot \theta}{r^4}V- \frac{\mu}{3 r^4 e^{2\Phi+\Lambda}}W_{,\theta \theta}+\frac{\mu}{r^4 e^{-2\Phi-\Lambda}}\biggl(\Phi^{\prime}-\frac{2}{r}\biggr)W_{\theta}.
\end{split}
\end{equation}

\bibliographystyle{mn2e}

\bibliography{paper_Colaiuda_Kokkotas_2011}

\begin{thebibliography}{}

\bibitem[\protect\citeauthoryear{{Andersson}, {Glampedakis} \&
  {Samuelsson}}{{Andersson} et~al.}{2009}]{2009MNRAS.396..894A}
{Andersson} N.,  {Glampedakis} K.,    {Samuelsson} L.,  2009, \mnras, 396, 894

\bibitem[\protect\citeauthoryear{{Braithwaite} \& {Nordlund}}{{Braithwaite} \&
  {Nordlund}}{2006}]{2006A&A...450.1077B}
{Braithwaite} J.,  {Nordlund} {\AA}.,  2006, \aap, 450, 1077

\bibitem[\protect\citeauthoryear{{Cerd{\'a}-Dur{\'a}n}, {Stergioulas} \&
  {Font}}{{Cerd{\'a}-Dur{\'a}n} et~al.}{2009}]{2009MNRAS.397.1607C}
{Cerd{\'a}-Dur{\'a}n} P.,  {Stergioulas} N.,    {Font} J.~A.,  2009, \mnras,
  397, 1607

\bibitem[\protect\citeauthoryear{{Ciolfi}, {Lander}, {Manca} \&
  {Rezzolla}}{{Ciolfi} et~al.}{2011}]{2011ApJ...736L...6C}
{Ciolfi} R.,  {Lander} S.~K.,  {Manca} G.~M.,    {Rezzolla} L.,  2011, \apjl,
  736, L6+

\bibitem[\protect\citeauthoryear{{Colaiuda}, {Beyer} \& {Kokkotas}}{{Colaiuda}
  et~al.}{2009}]{2009MNRAS.396.1441C}
{Colaiuda} A.,  {Beyer} H.,    {Kokkotas} K.~D.,  2009, \mnras, 396, 1441

\bibitem[\protect\citeauthoryear{{Colaiuda}, {Ferrari}, {Gualtieri} \&
  {Pons}}{{Colaiuda} et~al.}{2008}]{2008MNRAS.385.2080C}
{Colaiuda} A.,  {Ferrari} V.,  {Gualtieri} L.,    {Pons} J.~A.,  2008, \mnras,
  385, 2080

\bibitem[\protect\citeauthoryear{{Colaiuda} \& {Kokkotas}}{{Colaiuda} \&
  {Kokkotas}}{2011}]{2011MNRAS.414.3014C}
{Colaiuda} A.,  {Kokkotas} K.~D.,  2011, \mnras, 414, 3014

\bibitem[\protect\citeauthoryear{{Duncan}}{{Duncan}}{1998}]{1998ApJ...498L..45%
D}
{Duncan} R.~C.,  1998, \apjl, 498, L45+

\bibitem[\protect\citeauthoryear{{Gabler}, {Cerd{\'a} Dur{\'a}n}, {Font},
  {M{\"u}ller} \& {Stergioulas}}{{Gabler} et~al.}{2011}]{2011MNRAS.410L..37G}
{Gabler} M.,  {Cerd{\'a} Dur{\'a}n} P.,  {Font} J.~A.,  {M{\"u}ller} E.,
  {Stergioulas} N.,  2011, \mnras, 410, L37

\bibitem[\protect\citeauthoryear{{Gabler}, {Cerd{\'a}-Dur{\'a}n},
  {Stergioulas}, {Font} \& {M{\"u}ller}}{{Gabler}
  et~al.}{2012}]{2012MNRAS.421.2054G}
{Gabler} M.,  {Cerd{\'a}-Dur{\'a}n} P.,  {Stergioulas} N.,  {Font} J.~A.,
  {M{\"u}ller} E.,  2012, \mnras, 421, 2054

\bibitem[\protect\citeauthoryear{{Glampedakis}, {Andersson} \&
  {Samuelsson}}{{Glampedakis} et~al.}{2011}]{2011MNRAS.410..805G}
{Glampedakis} K.,  {Andersson} N.,    {Samuelsson} L.,  2011, \mnras, 410, 805

\bibitem[\protect\citeauthoryear{{Glampedakis}, {Samuelsson} \&
  {Andersson}}{{Glampedakis} et~al.}{2006}]{2006MNRAS.371L..74G}
{Glampedakis} K.,  {Samuelsson} L.,    {Andersson} N.,  2006, \mnras, 371, L74

\bibitem[\protect\citeauthoryear{{Hambaryan}, {Neuh{\"a}user} \&
  {Kokkotas}}{{Hambaryan} et~al.}{2011}]{2011A&A...528A..45H}
{Hambaryan} V.,  {Neuh{\"a}user} R.,    {Kokkotas} K.~D.,  2011, \aap, 528,
  A45+

\bibitem[\protect\citeauthoryear{{Haskell}, {Samuelsson}, {Glampedakis} \&
  {Andersson}}{{Haskell} et~al.}{2008}]{2008MNRAS.385..531H}
{Haskell} B.,  {Samuelsson} L.,  {Glampedakis} K.,    {Andersson} N.,  2008,
  \mnras, 385, 531

\bibitem[\protect\citeauthoryear{{Israel}, {Belloni}, {Stella}, {Rephaeli},
  {Gruber}, {Casella}, {Dall'Osso}, {Rea}, {Persic} \& {Rothschild}}{{Israel}
  et~al.}{2005}]{I2005}
{Israel} G.~L.,  {Belloni} T.,  {Stella} L.,  {Rephaeli} Y.,  {Gruber} D.~E.,
  {Casella} P.,  {Dall'Osso} S.,  {Rea} N.,  {Persic} M.,    {Rothschild}
  R.~E.,  2005, \apjl, 628, L53

\bibitem[\protect\citeauthoryear{{Kokkotas}, {Gaertig} \&
  {Colaiuda}}{{Kokkotas} et~al.}{2010}]{2010JPhCS.222a2031K}
{Kokkotas} K.~D.,  {Gaertig} E.,    {Colaiuda} A.,  2010, Journal of Physics
  Conference Series, 222, 012031

\bibitem[\protect\citeauthoryear{{Lander} \& {Jones}}{{Lander} \&
  {Jones}}{2011}]{2011MNRAS.412.1730L}
{Lander} S.~K.,  {Jones} D.~I.,  2011, \mnras, 412, 1730

\bibitem[\protect\citeauthoryear{{Lander}, {Jones} \& {Passamonti}}{{Lander}
  et~al.}{2010}]{2010MNRAS.405..318L}
{Lander} S.~K.,  {Jones} D.~I.,    {Passamonti} A.,  2010, \mnras, 405, 318

\bibitem[\protect\citeauthoryear{{Lasky}, {Zink}, {Kokkotas} \&
  {Glampedakis}}{{Lasky} et~al.}{2011}]{2011ApJ...735L..20L}
{Lasky} P.~D.,  {Zink} B.,  {Kokkotas} K.~D.,    {Glampedakis} K.,  2011,
  \apjl, 735, L20+

\bibitem[\protect\citeauthoryear{{Lee}}{{Lee}}{2008}]{2008MNRAS.385.2069L}
{Lee} U.,  2008, \mnras, 385, 2069

\bibitem[\protect\citeauthoryear{{Levin}}{{Levin}}{2006}]{2006MNRAS.368L..35L}
{Levin} Y.,  2006, \mnras, 368, L35

\bibitem[\protect\citeauthoryear{{Levin}}{{Levin}}{2007}]{2007MNRAS.377..159L}
{Levin} Y.,  2007, \mnras, 377, 159

\bibitem[\protect\citeauthoryear{{Levin} \& {van Hoven}}{{Levin} \& {van
  Hoven}}{2011}]{2011MNRAS.tmp.1405L}
{Levin} Y.,  {van Hoven} M.,  2011, \mnras, pp 1405--+

\bibitem[\protect\citeauthoryear{{Negele} \& {Vautherin}}{{Negele} \&
  {Vautherin}}{1973}]{1973NuPhA.207..298N}
{Negele} J.~W.,  {Vautherin} D.,  1973, Nuclear Physics A, 207, 298

\bibitem[\protect\citeauthoryear{{Pandharipande}}{{Pandharipande}}{1971}]{1971%
NuPhA.178..123P}
{Pandharipande} V.~R.,  1971, Nuclear Physics A, 178, 123

\bibitem[\protect\citeauthoryear{{Samuelsson} \& {Andersson}}{{Samuelsson} \&
  {Andersson}}{2007}]{SA2007}
{Samuelsson} L.,  {Andersson} N.,  2007, \mnras, 374, 256

\bibitem[\protect\citeauthoryear{{Sotani}}{{Sotani}}{2011}]{2011MNRAS.417L..70%
S}
{Sotani} H.,  2011, \mnras, 417, L70

\bibitem[\protect\citeauthoryear{{Sotani} \& {Kokkotas}}{{Sotani} \&
  {Kokkotas}}{2009}]{2009MNRAS.395.1163S}
{Sotani} H.,  {Kokkotas} K.~D.,  2009, \mnras, 395, 1163

\bibitem[\protect\citeauthoryear{{Sotani} \& {Kokkotas}}{{Sotani} \&
  {Kokkotas}}{2011}]{2011JPhCS.314a2081S}
{Sotani} H.,  {Kokkotas} K.~D.,  2011, Journal of Physics Conference Series,
  314, 012081

\bibitem[\protect\citeauthoryear{{Sotani}, {Kokkotas} \&
  {Stergioulas}}{{Sotani} et~al.}{2007}]{2007MNRAS.375..261S}
{Sotani} H.,  {Kokkotas} K.~D.,    {Stergioulas} N.,  2007, \mnras, 375, 261

\bibitem[\protect\citeauthoryear{{Sotani}, {Kokkotas} \&
  {Stergioulas}}{{Sotani} et~al.}{2008}]{2008MNRAS.385L...5S}
{Sotani} H.,  {Kokkotas} K.~D.,    {Stergioulas} N.,  2008, \mnras, 385, L5

\bibitem[\protect\citeauthoryear{{Strohmayer} \& {Watts}}{{Strohmayer} \&
  {Watts}}{2005}]{SW2005}
{Strohmayer} T.~E.,  {Watts} A.~L.,  2005, Ap. J. Lett., 632, L111

\bibitem[\protect\citeauthoryear{{van Hoven} \& {Levin}}{{van Hoven} \&
  {Levin}}{2011}]{2011MNRAS.410.1036H}
{van Hoven} M.,  {Levin} Y.,  2011, \mnras, 410, 1036

\bibitem[\protect\citeauthoryear{{van Hoven} \& {Levin}}{{van Hoven} \&
  {Levin}}{2012}]{2012MNRAS.420.3035V}
{van Hoven} M.,  {Levin} Y.,  2012, \mnras, 420, 3035

\bibitem[\protect\citeauthoryear{{Watts} \& {Strohmayer}}{{Watts} \&
  {Strohmayer}}{2006}]{Ws2006a}
{Watts} A.~L.,  {Strohmayer} T.~E.,  2006, \apjl, 637, L117

\bibitem[\protect\citeauthoryear{{Watts} \& {Strohmayer}}{{Watts} \&
  {Strohmayer}}{2007}]{2007AdSpR..40.1446W}
{Watts} A.~L.,  {Strohmayer} T.~E.,  2007, Advances in Space Research, 40, 1446

\bibitem[\protect\citeauthoryear{{Zink}, {Lasky} \& {Kokkotas}}{{Zink}
  et~al.}{2012}]{2012PhRvD..85b4030Z}
{Zink} B.,  {Lasky} P.~D.,    {Kokkotas} K.~D.,  2012, \prd, 85, 024030

\end{thebibliography}

\end{document}